# Twist-Tuned Magnonic Nanocavity Mode in a Trilayer Moiré Superlattice

Tianyu Yang[1], Gianluca Gubbiotti[2,*], Marco Madami,[3] Haiming Yu[1], Jilei Chen[4,*]

[1]Fert Beijing Institute, MIIT Key Laboratory of Spintronics, School of Integrated Circuit Science and Engineering, Beihang University, Beijing, China

[2] CNR-Istituto Officina dei Materiali (IOM), 06123 Perugia, Italy

[3] Dipartimento di Fisica e Geologia, Università di Perugia, Perugia I-06123, Italy

[4]International Quantum Academy, Shenzhen, China

E-mail address：

* gubbiotti@iom.cnr.it

* chenjilei@iqasz.cn

**ABSTRACT** The concept of moiré superlattices has recently been introduced into the field of magnonics, enabling unprecedented control over spin-wave propagation and confinement in nanoscale magnonic devices. In this work, we report a numerical investigation on the nanocavity in a trilayer magnetic moiré superlattice structure consisting of antidot lattices. By tuning the middle layer twist angle, high tunability of the magnonic band structure can be achieved with characteristic flat bands and the corresponding nanocavity mode formation in outer layers. At an optimal twist angle of 3°, excitation at the flat band frequency yields nanocavity mode with linewidth of 175 nm. In contrast to its bilayer counterpart, the trilayer magnonic moiré superlattice exhibits antiphase nanocavity modes in the outer layers while showing no nanocavity formation in the middle layer. Our study indicates that the switching and distribution of the nanocavity modes can be governed by tuning the middle layer twist angle with a strong magnon intensity confinement. The trilayer magnonic moiré structure holds a distinct advantage in tunability, which opens up new avenues for the design of future moiré magnonic devices.

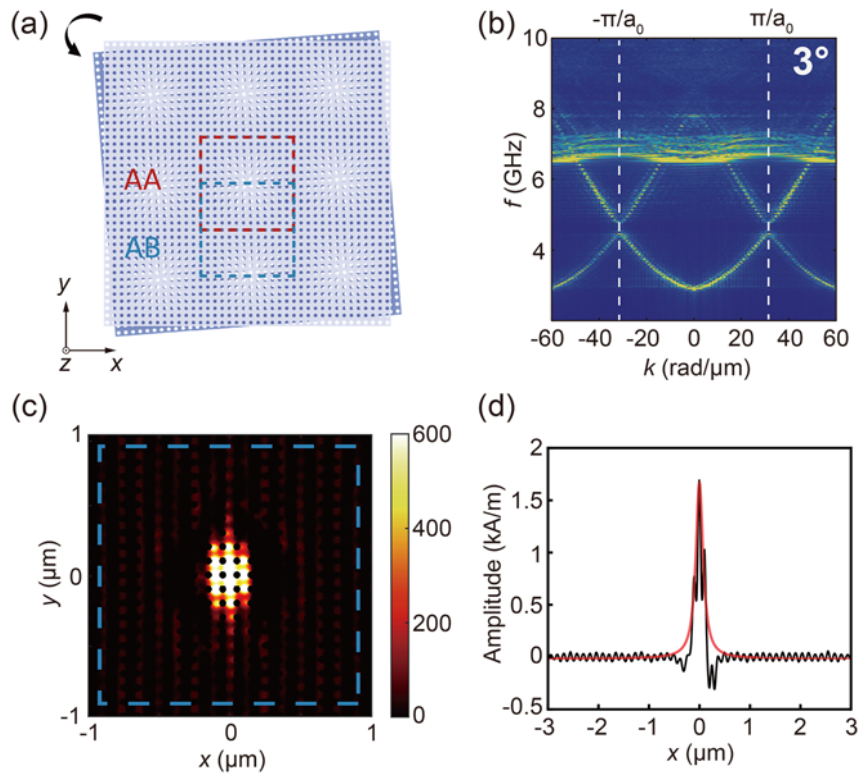

## INTRODUCTION

A moiré superlattice can be formed in vertically stacked periodic two-dimensional (2D) layers with a small angular or lattice mismatch between the two monolayers. Recently, it has achieved numerous exotic results in moiré twisted bilayer graphene[1–6] and transition metal dichalcogenide,[7–9] such as unconventional superconductivity,[10] Mott insulators[11] and a variety of correlated quantum phenomena.[12–16] At a "magic angle" of 1.1°, electronic structures are reconstructed and host nearly flat bands in twisted bilayer graphene. Therefore, adjusting the twist angle has shown potential strategy to tuning and manipulate electronic properties.[17,18] In addition to twisted bilayer, the family of twisted moiré superlattice has been expanded to multilayer systems, which consist of three or more layers with a specific rotational structure. Specifically, the moiré pattern in trilayer structure is controlled by two independent twisting angles and their combinations creates a vast and complex parameter space with additional degrees of freedom to control the electronic structure. [19–24]

Magnons, the quanta of spin waves (SWs),[25–27] are collective spin excitations which can propagate coherently in ferromagnetic magnetic materials free of Joule heating because of the absence of electron transport.[28,29] The emerging field of magnonics is ideally suited to serve as the nanoscale platform for analog computing. Essentially, magnonics can be regarded as interference-based, nonlinear, and short-wavelength optics implemented with SWs, which intrinsically enable low-power operation. It holds strong potential for enabling low power devices[30] and logic circuits[31] that can extend beyond conventional CMOS technology.[32]

Moiré physics has recently been introduced toward the field of magnonics as an alternative approach for controlling and tuning the SW propagation properties through band structure engineering.[33–37] The magnon flatband and the resulting nanocavity mode formed at specific frequencies has been reported in twisted bilayer magnonic crystals through micromagnetic simulation.[38,39] Wang *et al.* observed edge and cavity modes in an artificial twisted moiré superlattices consisting of two sets of antidot lattices etched into one at the same yttrium iron garnet (YIG) thin film.[40] In addition, the exploration of magnons in twisted magnetic van der Waals (vdW) moiré structures[41–44] is also a highlight for the development of magnonic moiré devices, such as the discovery that stacking domain walls in bilayer twisted $CrI_3$ can host one-dimensional (1D) magnon channels with lower energy than bulk magnons.[45] So far, most explored researches for magnon in moiré superlattice are focusing on bilayer moiré magnonic devices. In contrast, magnonic in twisted trilayer moiré superlattice have received significantly less attention despite their improved tunability, encompassing three independently rotatable layers and two interfaces with adjustable interlayer exchange coupling.

In this paper, we study by micromagnetic simulations the properties of magnons in a trilayer twisted magnetic moiré superlattice. Through microwave excitation on the bottom antidot layer only (asymmetric excitation), magnons were excited in the magnetic trilayer moiré superlattice with a twisted middle layer. The subsequent emergence of flat bands in SW band structures of bottom and top layers led to the formation of nanocavity modes. It is observed that the middle layer twist angle provides full control over the switching, spatial distribution, and 180° phase shift of these out-of-phase nanocavity modes, suggesting its potential for future vertical magnon-based nanoscale transistors, phase shifters and 3D signal-transmission devices.[46,47] Furthermore, the higher frequency modes located above the primary flat band were found to generate additional nanocavity modes. A card-deck-like trilayer structure formed by twisting both the bottom and top layers with respect to the central layer is also investigated.[48,49] These two mismatched moiré patterns result in a flat band with a narrower linewidth, which corresponds to a relatively weaker nanocavity mode. The trilayer magnetic moiré superlattice structure offers greater tunability compared to its bilayer counterpart, thereby presenting broader prospects for the development of novel magnonic devices.

**RESULTS AND DISCUSSION**

We consider a trilayer moiré superlattice system comprising three YIG layers, which serves as a prototypical material platform for investigating coherent SW propagation due to its inherently ultra-low Gilbert damping.[50–52] Each layer is patterned into a square antidot array[53–55] with periodicity of $a_0$ = 100 nm and antidot diameter $d$ = 50 nm, as shown in Figure 1a. The single layer thicknesss is fixed at 2 nm with adjustable interlayer exchange coupling between each two layers (Figure 1b). The magnetization and damping are set as $M_S$ = 140 kA/m and $\alpha$ = 1×10$^{-4}$. The micromagnetic simulation is conducted in a cubic spatial domain of dimensions 6 $\mu$m × 6 $\mu$m × 6 nm and adjacent layers maintain interfacial contact to facilitate direct interlayer exchange coupling. A bias magnetic field $H$ = 50 mT is applied along the $y$-direction while an out-of-plane sinusoidal microwave pulse is applied in the central stripline region of the bottom layer, asymmetric excitation, to excite magnons propagating along the $x$-direction, resulting in a Damon-Eshbach configuration.[56,57] The SW dispersion relations are calculated by performing a fast Fourier transform (FFT) on the magnetization component $m_x$ along the center of the moiré unit cell. Figure 1c-f present the magnonic band structures of different interlayer exchange coupling without any layer twist. In Figure 1c, it can be observed that several sets of bands are overlapping in the trilayer stack without interlayer exchange interaction, and one band have a slightly higher energy. When a finite interlayer exchange coupling is applied between the antidot lattices, some of the bands are upshifted in frequency while some others remain unchanged. These bands are labeled as Band 1, Band 2, and Band 3 according to their respective band bottom in

Figure 1d. Notably, Bands 2 and 3 exhibit an opposite concavity compared to the other bands, which remain largely unaffected by variations in the interlayer exchange coupling. Under conditions of weak exchange coupling, the three band sets remain clearly visible with complete structures (see Supporting Information Figure S1). As the interlayer exchange increase, both Band 2 and Band 3 are observed to be upshifted in frequency, and Band 3 displays a larger shift compared to Band 2, whereas the position of Band 1 remains unchanged. Band 3 corresponds to the magnonic band of the middle layer, which is also the one with higher energy in Figure 1c (marked in white arrow). This increased energy is probably ascribed to the dipolar interaction from both adjacent layers in the absence of interlayer exchange coupling. Band 2 corresponds to the magnonic bands of the top and bottom layers. Since it experiences interlayer exchange coupling from only one side, its energy shift is hence smaller than that of Band 3 when the exchange interaction increases. This interpretation will be further confirmed by the each-layer dispersion relations extracted subsequently. In Figure 1e and f, the interlayer exchange interaction progressively increases, both Band 2 and Band 3 continue to shift upward, with Band 3 eventually shifting beyond the 10 GHz range of the spectrum.

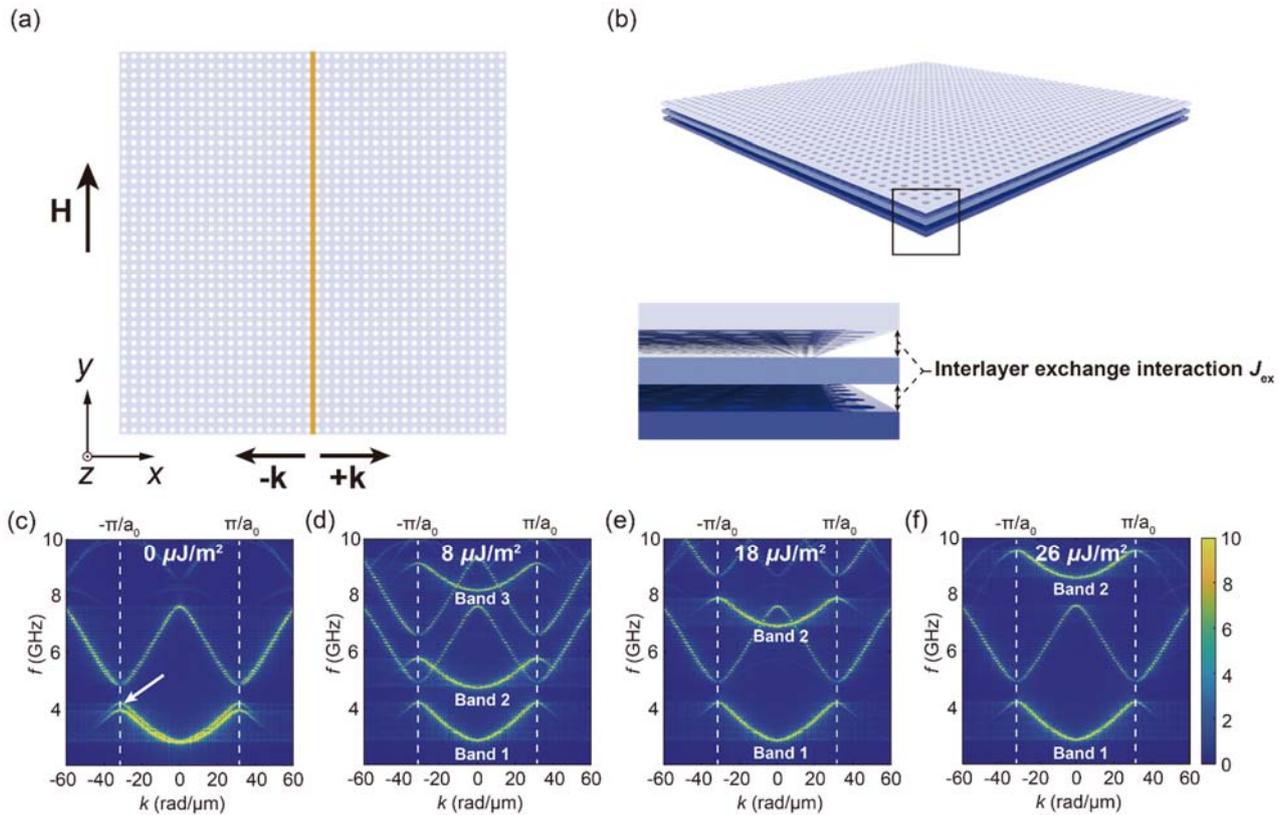

**Figure 1. SW excitation in trilayer YIG magnonic lattice. (a) Illustrative diagram of trilayer magnonic lattice. An external magnetic field *H* is applied along the y-axis. A sinusoidal microwave field is applied in the gold stripline region to excite SW along the x-axis. (b) Side view and cross-section of the trilayer magnonic lattice. The interlayer exchange interaction $J_{ex}$ is indicated between each two layers. (c)-(f) Magnonic band structures with the interlayer**

**exchange coupling of 0 $\mu J/m^2$, 8 $\mu J/m^2$, 18 $\mu J/m^2$ and 26 $\mu J/m^2$. The white arrow in Figure 1(c) mark a doublet of branches resulting from the interlayer dipolar interaction. The white dashed line indicates the Brillouin zone boundaries.**

To investigate the band characteristics of the trilayers moiré magnonic crystal, we varied the twist angle of the middle layer to produce a moiré pattern with a moiré unit cell size governed by $a_m = a_0/\theta$ [58] at a fixed interlayer exchange coupling $J_{ex} = 26$ $\mu J/m^2$. Figures 2a and 2b show the top and side views of the trilayer moiré superlattice. The centre of the moiré superlattice unit cell (commensurate AA region) and the edge of the unit cell (incommensurate AB region) are marked by red and blue dashed boxes respectively.[59] Figures 2c-2f illustrate the evolution of SW band structure as the twist angle of the middle layer increases from 0° to 6°. It is observed that Band 2, which corresponds to the magnon bands of the bottom and top layers, decreases in frequency and gradually evolves into multiple flat bands, as shown in Figure 2c. Furthermore, the flat band reaches its optimal flatness at a twist angle of 3° for middle layer. Moreover, unlike the flat band structures previously reported in bilayer magnonic moiré superlattices,[38] the trilayer system exhibits additional higher-order above the main flat band, as shown in Figure 2e. The spatial distribution of magnetic dynamical component $m_x$ is extracted to analyze the SW propagation at the flat band frequency of 6.5 GHz. As expected, the SW propagation in an untwisted trilayer structure matches exactly that of a conventional waveguide, as presented in Figure 2g. However, as shown in Figure 2h-2j, when a twist angle is introduced in the middle layer, the SW spatial profile exhibits a strong localization giving rise to nanocavity modes at the center of AB stacking region owing to the zero group velocity of magnons at the flat band. The snapshot taken at 3° twist angle displays the most well-defined nanocavity modes, which is consistent with the high quality of its corresponding flat band structure. The mode profiles for $\theta = 1°$ exhibits an arrowhead-like shape whereas at $\theta = 6°$, it becomes highly diffuse and barely distinguishable.

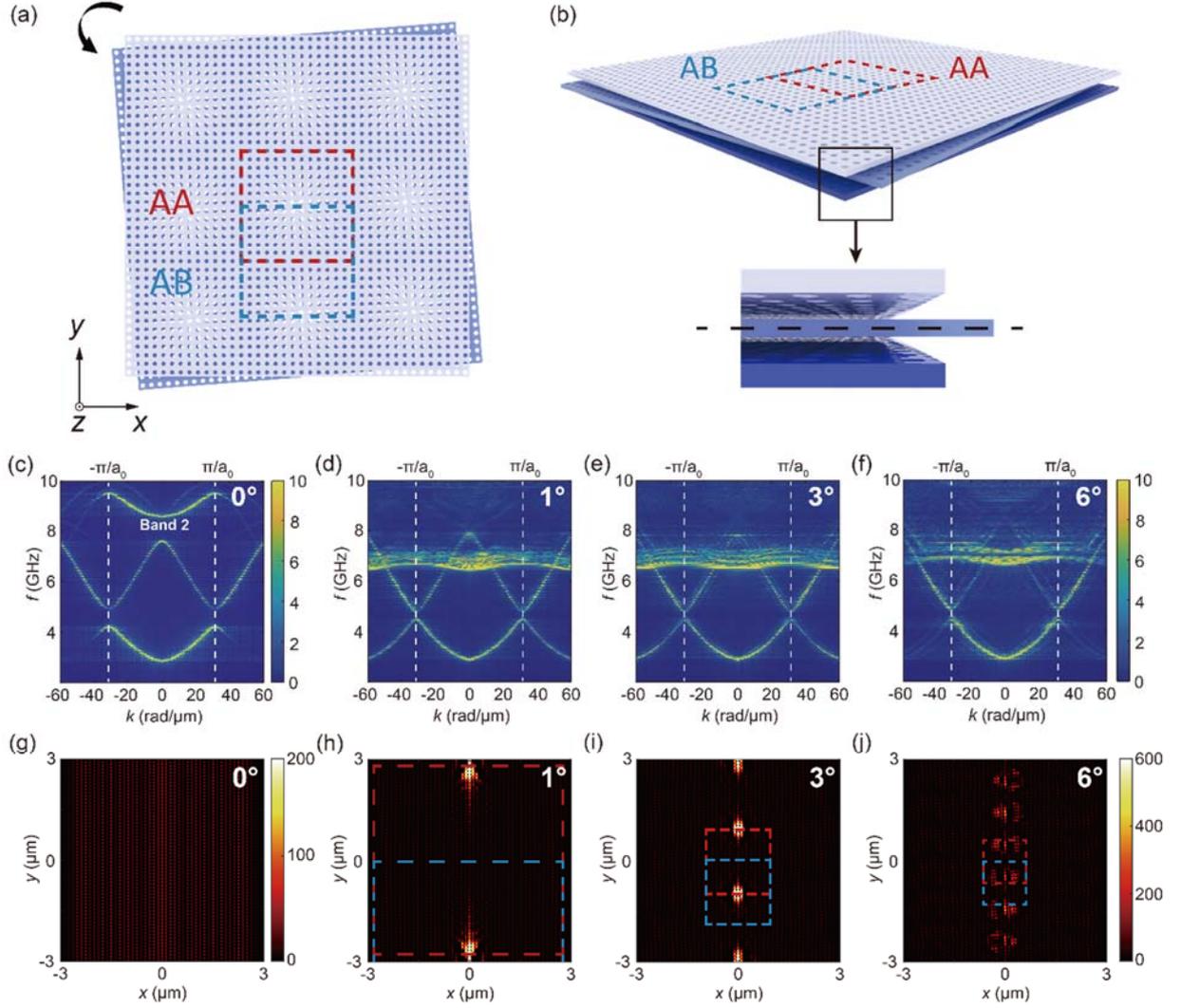

**Figure 2. SW excitation in trilayer moiré superlattice.** (a) Schematic of trilayer moiré superlattice with a twist angle of 3° in the middle layer. The AA and AB stacking regions of moiré superlattice are indicated by red and blue dashed lines. (b) Side view of the magnetic moiré superlattice with a twisted middle layer with a magnified view, the black dashed line denotes the axis of symmetry. (c)-(f) Magnonic band structures calculated in the middle of AB stacking region with different twist angles of 0°, 1°, 3° and 6°. The interlayer exchange interaction is fixed at 26 $\mu J/m^2$. (g)-(j) The mode profiles snapshots of the magnetization dynamics $m_x$ in the bottom layer at $t$ = 4 ns with excitation frequency of 6.5 GHz and twist angles of 0°, 1°, 3° and 6°. The red and blue dashed square represent the corresponding AA and AB stacking regions in the mode profiles snapshots.

Therefore, the middle layer twist angle of 3° was selected for an in-depth investigation of the nanocavity modes in trilayer moiré magnonic superlattice and considered as the "magic" angle for our trilayer structure. Figure 2e displays its SW dispersion of this configuration with distinct flat band structure at 6.5 GHz, spanning a wavevector range from -17.5 rad/μm to 17.5 rad/μm. Figure 3a provides an enlarged 1 μm × 1 μm image focusing on the nanocavity region for the bottom layer. The magnons are effectively confined at the center of the AB stacking region forming a nanocavity. Furthermore, the mode profile exhibit an approximate mirror symmetry[60–65] about the horizontal axis, in accordance with the geometric symmetry of the trilayer system with a rotated middle layer, as presented in the magnified view of Figure 2b. A linecut of magnetization component $m_x$ (black curve) extracted from the center of AB region ($y = 0$) is presented in Figure 3b. A pronounced enhancement of the magnon intensity at the center ($x = 0$) is observed, directly evidencing the formation of a nanocavity mode. Beyond the nanocavity region, SW maintain stable propagation characteristics in both intensity and wavelength, though their signal strength is significantly weaker. Performing a Lorentzian fit on this linecut yields a linewidth of 174.9 nm. The corresponding wavevector range is calculated as $\Delta k = \frac{2\pi}{\Delta x} = 35.9$ rad/μm, which agrees with the value extracted from Figure 2e.

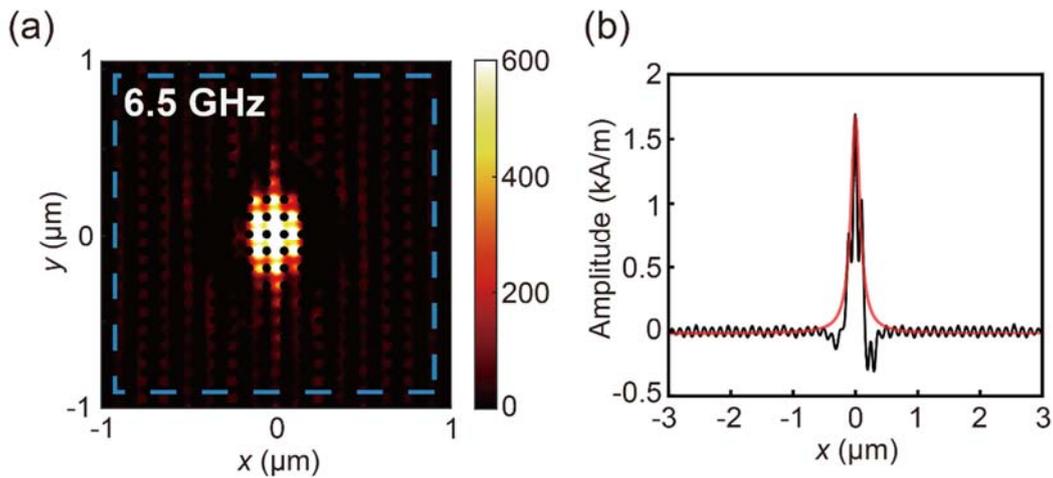

**Figure 3. Nanocavity mode in trilayer moiré superlattice. (a) The mode profile snapshot of the magnetization dynamics $m_x$ at $t = 4$ ns in the bottom layer with the excitation frequency of 6.5 GHz. The blue dashed square represents the moiré unit cell with AB stacking region. (b) A linecut of the magnetization intensity at y = 0 in (a). The Lorentz fitting (red curve) yields a linewidth of Δx = 174.9 nm.**

To gain further insight into the behavior of nanocavity mode in trilayer moiré magnonic lattice, we perform two-dimensional Fourier transform for each layer thus exploring the band structure for each of them. The bottom and top layers show nearly identical SW dispersion relations featuring a flat band at 6.5 GHz (Figure 4a and 4c). In contrast, no flat band appears in the middle layer. This is expected because the flat band is formed by Band 2, which corresponds to the magnonic modes of the bottom and top layers. Since Band 2 is absent in the middle layer, the flat band does not form there. This feature is further confirmed by the band structure of each single layer without twist angle (Supporting Information Figure S2). The extraction of the dynamic magnetization component $m_x$ for each layer at flat band (6.5 GHz) allows for an investigation to the link between band structures and nanocavity properties. The first key observation is the disappearance of nanocavity mode in the middle layer in Figure 4e, which is consistent with the absence of a flat band in its corresponding band structure. Then, the nanocavity modes with out-of-phase oscillation precession are presented in the AB stacking region of bottom and top layers result from their flat band structure at 6.5 GHz, as presented in Figure 4d and 4f. This observation indicates that although the bottom and top layers host nearly identical flat bands in their magnonic spectra, these flat bands represent precessions with opposite phases due to the twist. Simultaneously, the interactions of the bottom and top layers cancel within the middle layer due to their out-of-phase spin precession, resulting in the disappearance of Band 2. In addition, the twist induced magnonic lattice mismatch leads to a different variations in SW amplitude along AA and AB stacking regions (see Supporting Information Figure S3).

It is worth noticing that all these results are obtained for the case of asymmetric excitation where the magnetic field pulse is applied to the bottom layer, only. The phase of nanocavity modes in the bottom and top layers can be reversed by switching the excitation from bottom to top layer. Whereas, nanocavity formation is entirely absent when the excitation is applied to the middle layer or to all three layers simultaneously. Thus, asymmetric excitation is required for nanocavity mode formation (Supporting Information Figure S4). The vertical operation of the nanocavity mode in trilayer moiré superlattice could function as a nanoscale transistor: When a microwave signal is injected into the bottom layer (transistor source), the nanocavity mode can be switched on/off by rotating the middle layer (transistor gate) enabling the transmission of nanocavity modes with opposite phases to top layer (transistor drain), while the middle layer itself remains free of nanocavity formation. Furthermore, a rapid 180° phase shift between the bottom and top layers can be achieved within 0.1 ns while maintaining the signal amplitude and shape (see Supporting Information Figure S5), demonstrating its potential for implementing a high-speed vertical phase shifter in the trilayer moiré magnonic lattice. Moreover, the spatial distribution of the nanocavity mode can be governed by the moiré period, which is adjusted via the middle layer twist angle, as evidenced by the mode profiles at twist angle of 1° and 6°, as shown in Supporting Information Figure S6.

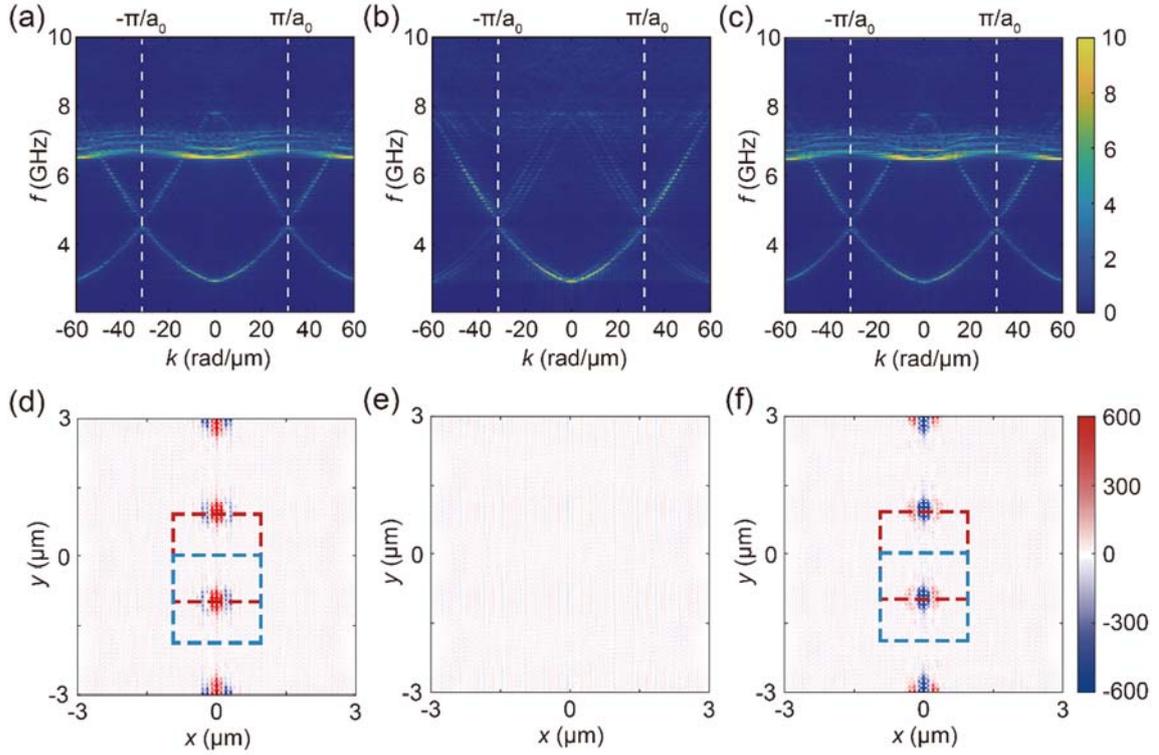

**Figure 4. The behavior of nanocavity mode in each layer of trilayer supperlattice with a twist angle of 3° in the middle layer. (a)-(c) The SW band structures of bottom layer, middle layer and top layer. (d)-(e) The mode profiles at *t* = 4 ns with the excitation frequency of 6.5 GHz extracted from bottom, middle and top layers. The colorbar range and color map were optimized to best visualize the phase signature of the nanocavity in each layer. The red and blue in the colormaps represent opposite phase of the magnetization dynamics $m_x$.**

A closer investigation of the flat band region around $k = 0$ (Figure 5a) shows higher-order flat bands at frequencies above 6.5 GHz, suggesting a more complex mode structure. A magnified spectrum above the 6.5 GHz flat band uncovers two high-order flat bands at 6.75 GHz and 6.97 GHz marked with black dotted line, as shown in Figure 5b. As further evidence for the higher-order modes, the magnonic density of states (DOS) from the region outlined in Figure 5a was analyzed. The resulting DOS spectrum shows several subsidiary peaks above the main peak at 6.5 GHz, providing direct evidence for the higher order flat bands. Subsequently, SWs were excited at the frequencies of the higher-order modes. These modes are also found to generate nanocavity modes at a same time step (see Figure 5d-f). Differing from the fundamental flat band mode at 6.5 GHz that is spatially confined to the center, the higher-order modes reveal a hollow spatial profile, as clearly displayed in the Supporting Information Figure S7. These confined modes similar to edge modes found in certain finite-size photonic crystals ,which show that in confined structure there exist isolated instead of a continuum of modes.[66] In AB stacking region, the moiré magnetic nanostructure engineers the internal field, resulting in separated magnon modes that are localized in a circular region at higher frequency.

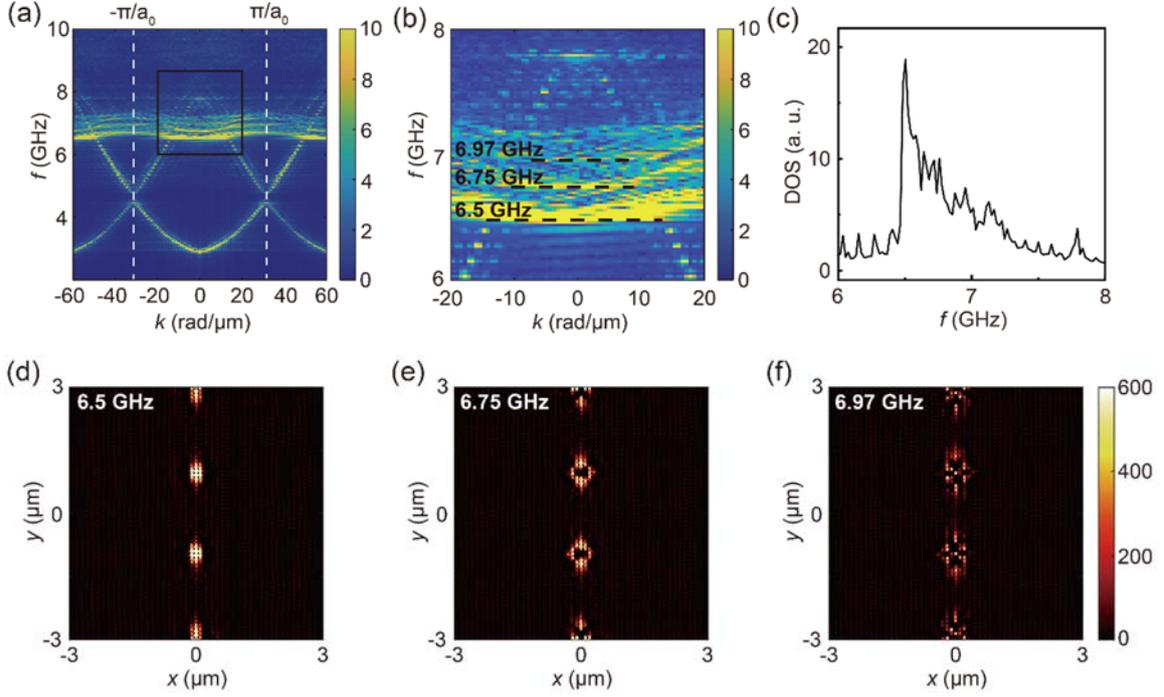

**Figure 5. High order modes above the flat band. (a) The band structure with a twist angle of 3° with the interlayer exchange interaction of 26 $\mu J/m^2$ in the middle layer. The black square indicates the area of SW high order modes. (b) SW dispersion near the flat band. High order modes are indicated by black lines at 6.5 GHz, 6.75 GHz and 6.97 GHz. (c) Magnon density of states (DOS) depending on the excitation frequency. (d)-(f) The mode profiles of the magnetization dynamics $m_x$ in bottom layer at $t = 4$ ns with the excitation frequencies of 6.5 GHz, 6.75 GHz and 6.97 GHz corresponding to the three modes in (b).**

The mirror-symmetric structure, achieved by twisting only the middle layer, is characterized by relative twist angles between the bottom-middle and middle-top layer pairs that are equal in magnitude but opposite in sign. Assigning the same sign to both relative twist angles could lead to different physical properties. In this respect, we focus on a specific case where the bottom layer is rotated by +3° and the top layer by -3° while the middle layer is fixed, forming a card-deck-like trilayer structure, as presented in Figure 6a and b. Figure 6c displays the SW dispersion relation obtained from a 2D FFT along the center of its AB region which also shows a flat band feature at 6.38 GHz. However, this flat band is markedly weaker and has a significantly reduced linewidth compared to the one found in the band structure with a 3° middle layer twist. As shown in the corresponding snapshot of the dynamic magnetization (Figure. 6d), a low amplitude nanocavity mode is present in the center of the AB region highlighted by the blue dashed box. The nanocavity mode is slightly elongated and rotated with respect to the magnetic field direction losing its symmetric shape observed in the case where only the middle layer is twisted. The configuration with only a twisted middle layer supports a higher quality flat band and a stronger nanocavity mode than the card-deck-like structure originating from the precise alignment of the top and bottom layers, which leads to a single moiré periodicity. In contrast, the card-deck-like trilayer structure hosts two misligned moiré periodicities within the trilayer magnonic crystal resulting in a more complex moiré-of-moiré pattern.[67] This complexity likely leads to mutual interference between these two moiré periods, which consequently weakens the intensity of nanocavity mode. The presence of two tunable interfaces substantially expands the research direction for trilayer magnonic moiré crystals.

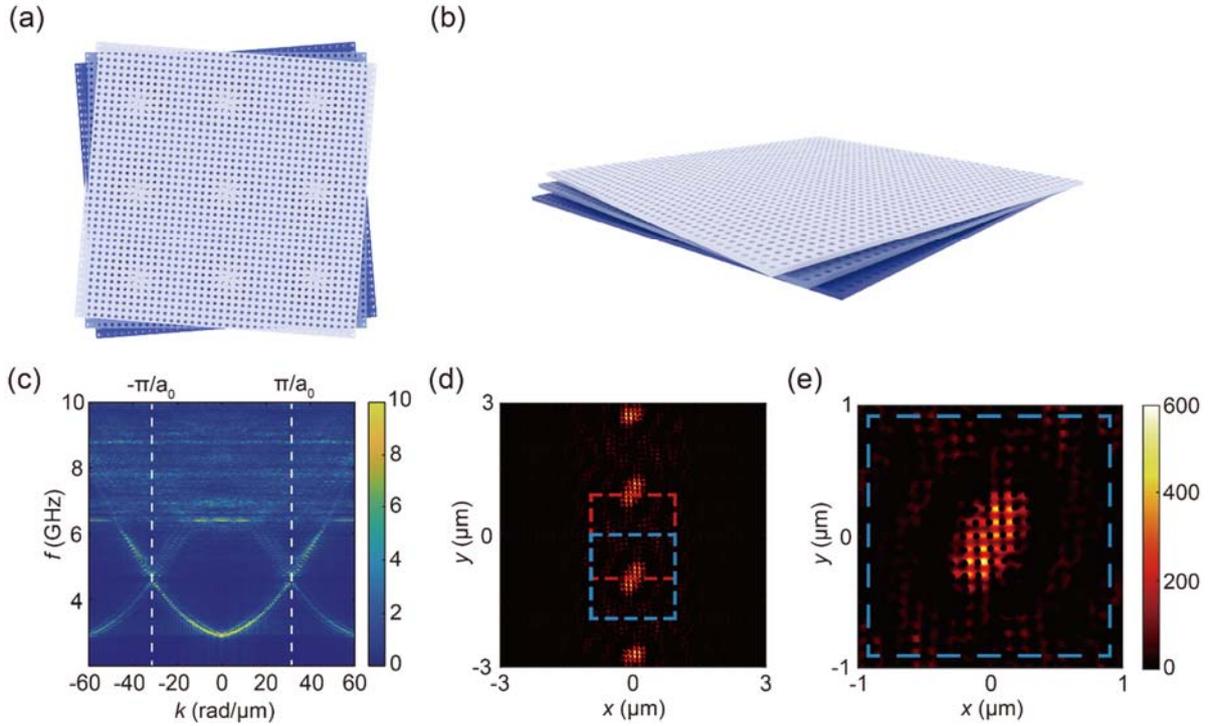

**Figure 6. SW excitation in card-deck-like trilayer moiré superlattice. (a) A Card-deck-like trilayer structure with a twist angle of +3° in the bottom layer and -3° in the top layer. (b) Side view of the trilayer card-deck-like trilayer moiré superlattice. (c) SW band structure of card-deck-like trilayer with the interlayer exchange interaction of 26 $\mu J/m^2$ between each two layers. (d) The magnetization dynamics $m_x$ snapshot in bottom layer at $t$ = 3.95 ns with an excitation frequency of 6.38 GHz.**

## CONCLUSIONS

To summarize, we have investigated the spin-wave band structure in a trilayer magnonic moiré superlattice using micromagnetic simulations. Twisting the middle layer generates a unique flat band in the magnonic spectrum, associated with two antiphase nanocavity modes localized in the AB regions of the bottom and top layers. By tuning the twist angle, both the switching behavior and the spatial distribution of these nanocavity modes can be controlled, while also enabling a 180° phase shift to be achieved on a sub-nanosecond timescale. Higher-order modes located above the primary flat band give rise to additional nanocavity edge modes that exhibit a distinctive spatial profile. Overall, the trilayer magnetic moiré structure offers greater tunability than its bilayer counterpart, opening new routes for the development of advanced moiré magnonic devices.

## METHODS

**Micromagnetic simulations**

The micromagnetic simulations were performed using the GPU-accelerated finite-difference program MuMax3[68]. The magnetization dynamics of the entire system were obtained by numerically solving the Landau-Lifshitz-Gilbert (LLG) equation in each cell:

$$\frac{d\boldsymbol{M}}{dt} = -\gamma(\boldsymbol{M} \times \boldsymbol{H}) + \frac{\alpha}{|\boldsymbol{M}|}\left(\boldsymbol{M} \times \frac{d\boldsymbol{M}}{dt}\right),$$

where $\boldsymbol{M}$ is the magnetization vector, $\boldsymbol{H}$ is the effective field acting on the magnetization vector, $\gamma$ is the gyromagnetic ratio and $\alpha$ is the Gilbert damping. The entire system was initially magnetized uniformly along the y-direction. To obtain the magnonic band structure, a sinusoidal function magnetic field pulse along the z-direction, given by $H_{\text{rf}} = H_0 \sin(2\pi f_0 t)/2\pi f_0 t$ was then applied to the central stripline region to excite SWs, where $H_0$ = 2 mT and $f_0$ = 10 GHz. The total simulation time was set to 40 ns with a time step of 50 ps. The SW mode profiles were characterized using single-frequency excitation $H_{\text{rf}} = H_0 \sin(2\pi f_0 t)$, where $f_0$ was set to the flat band frequency. This trilayer simulation system allows for the excitation antenna to be positioned on any chosen layer.

## AUTHOR INFORMATION

**Corresponding Author**

* gubbiotti@iom.cnr.it

* chenjilei@iqasz.cn

**ORCID**

**Author Contributions**

J.C. and G.G. conceived and designed the project. T.Y. conducted the micromagnetic simulations. T.Y., J.C. and G.G. analyzed and interpreted the data. T.Y., J.C. and G.G. wrote the manuscript. All authors commented on the manuscript.

**Notes**

The authors declare no competing financial interest.

## ACKNOWLEDGMENTS

This work was supported by the National Key Research and Development Program of China under Grant No. 2022YFA1402801. G. Gubbiotti and H. Yu acknowledge the financial support of the bilateral agreement CNR/National Natural Science Foundation (CHINA), under project "Spin-orbit interaction based on topological insulator/ferromagnet heterostructures" (2024-2025). G. G. also acknowledge the European Union— Next Generation EU under the Italian Ministry of University and Research (MUR) National Innovation Ecosystem grant ECS00000041—VITALITY. CUP: B43C22000470005.

Supporting Information Available: SW excitation with very weak interlayer exchange, SW band structures in each layer of trilayer YIG magnonic lattice without twist angle, variations of SW amplitude along AA and AB stacking region, symmetric and asymmetric SW excitation in trilayer moiré supperlattice, time-dependent $m_x$ at the center of AB stacking region, control of nanocavity mode distribution though twist middle layer, mode profiles of high order modes.

## REFERENCES

(1) Inbar, A.; Birkbeck, J.; Xiao, J.; Taniguchi, T.; Watanabe, K.; Yan, B.; Oreg, Y.; Stern, A.; Berg, E.; Ilani, S. The Quantum Twisting Microscope. *Nature* **2023**, *614* (7949), 682–687. https://doi.org/10.1038/s41586-022-05685-y.
(2) Lopes dos Santos, J. M. B. Graphene Bilayer with a Twist: Electronic Structure. *Phys. Rev. Lett.* **2007**, *99* (25). https://doi.org/10.1103/PhysRevLett.99.256802.
(3) Lu, C.-C.; Lin, Y.-C.; Liu, Z.; Yeh, C.-H.; Suenaga, K.; Chiu, P.-W. Twisting Bilayer Graphene Superlattices. *ACS Nano* **2013**, *7* (3), 2587–2594. https://doi.org/10.1021/nn3059828.
(4) Huang, S.; Kim, K.; Efimkin, D. K.; Lovorn, T.; Taniguchi, T.; Watanabe, K.; MacDonald, A. H.; Tutuc, E.; LeRoy, B. J. Topologically Protected Helical States in Minimally Twisted Bilayer Graphene. *Phys. Rev. Lett.* **2018**, *121* (3), 037702. https://doi.org/10.1103/PhysRevLett.121.037702.
(5) Rickhaus, P.; Wallbank, J.; Slizovskiy, S.; Pisoni, R.; Overweg, H.; Lee, Y.; Eich, M.; Liu, M.-H.; Watanabe, K.; Taniguchi, T.; Ihn, T.; Ensslin, K. Transport Through a Network of Topological Channels in Twisted Bilayer Graphene. *Nano Lett.* **2018**, *18* (11), 6725–6730. https://doi.org/10.1021/acs.nanolett.8b02387.
(6) Yoo, H.; Engelke, R.; Carr, S.; Fang, S.; Zhang, K.; Cazeaux, P.; Sung, S. H.; Hovden, R.; Tsen, A. W.; Taniguchi, T.; Watanabe, K.; Yi, G.-C.; Kim, M.; Luskin, M.; Tadmor, E. B.; Kaxiras, E.; Kim, P. Atomic and Electronic Reconstruction at the van Der Waals Interface in Twisted Bilayer Graphene. *Nat. Mater.* **2019**, *18* (5), 448–453. https://doi.org/10.1038/s41563-019-0346-z.
(7) Devakul, T.; Crépel, V.; Zhang, Y.; Fu, L. Magic in Twisted Transition Metal Dichalcogenide Bilayers. *Nat Commun* **2021**, *12* (1), 6730. https://doi.org/10.1038/s41467-021-27042-9.
(8) Wang, L.; Shih, E.-M.; Ghiotto, A.; Xian, L.; Rhodes, D. A.; Tan, C.; Claassen, M.; Kennes, D. M.; Bai, Y.; Kim, B.; Watanabe, K.; Taniguchi, T.; Zhu, X.; Hone, J.; Rubio, A.; Pasupathy, A. N.; Dean, C. R. Correlated Electronic Phases in Twisted Bilayer Transition Metal Dichalcogenides. *Nat. Mater.* **2020**, *19* (8), 861–866. https://doi.org/10.1038/s41563-020-0708-6.

(9) Wu, F.; Lovorn, T.; Tutuc, E.; Martin, I.; MacDonald, A. H. Topological Insulators in Twisted Transition Metal Dichalcogenide Homobilayers. *Phys. Rev. Lett.* **2019**, *122* (8), 086402. https://doi.org/10.1103/PhysRevLett.122.086402.

(10) Cao, Y.; Fatemi, V.; Fang, S.; Watanabe, K.; Taniguchi, T.; Kaxiras, E.; Jarillo-Herrero, P. Unconventional Superconductivity in Magic-Angle Graphene Superlattices. *Nature* **2018**, *556* (7699), 43–50. https://doi.org/10.1038/nature26160.

(11) Cao, Y.; Fatemi, V.; Demir, A.; Fang, S.; Tomarken, S. L.; Luo, J. Y.; Sanchez-Yamagishi, J. D.; Watanabe, K.; Taniguchi, T.; Kaxiras, E.; Ashoori, R. C.; Jarillo-Herrero, P. Correlated Insulator Behaviour at Half-Filling in Magic-Angle Graphene Superlattices. *Nature* **2018**, *556* (7699), 80–84. https://doi.org/10.1038/nature26154.

(12) Polshyn, H.; Yankowitz, M.; Chen, S.; Zhang, Y.; Watanabe, K.; Taniguchi, T.; Dean, C. R.; Young, A. F. Large Linear-in-Temperature Resistivity in Twisted Bilayer Graphene. *Nat. Phys.* **2019**, *15* (10), 1011–1016. https://doi.org/10.1038/s41567-019-0596-3.

(13) Choi, Y.; Kemmer, J.; Peng, Y.; Thomson, A.; Arora, H.; Polski, R.; Zhang, Y.; Ren, H.; Alicea, J.; Refael, G.; von Oppen, F.; Watanabe, K.; Taniguchi, T.; Nadj-Perge, S. Electronic Correlations in Twisted Bilayer Graphene near the Magic Angle. *Nat. Phys.* **2019**, *15* (11), 1174–1180. https://doi.org/10.1038/s41567-019-0606-5.

(14) Xie, Y.; Lian, B.; Jäck, B.; Liu, X.; Chiu, C.-L.; Watanabe, K.; Taniguchi, T.; Bernevig, B. A.; Yazdani, A. Spectroscopic Signatures of Many-Body Correlations in Magic-Angle Twisted Bilayer Graphene. *Nature* **2019**, *572* (7767), 101–105. https://doi.org/10.1038/s41586-019-1422-x.

(15) Zondiner, U.; Rozen, A.; Rodan-Legrain, D.; Cao, Y.; Queiroz, R.; Taniguchi, T.; Watanabe, K.; Oreg, Y.; von Oppen, F.; Stern, A.; Berg, E.; Jarillo-Herrero, P.; Ilani, S. Cascade of Phase Transitions and Dirac Revivals in Magic-Angle Graphene. *Nature* **2020**, *582* (7811), 203–208. https://doi.org/10.1038/s41586-020-2373-y.

(16) Wong, D.; Nuckolls, K. P.; Oh, M.; Lian, B.; Xie, Y.; Jeon, S.; Watanabe, K.; Taniguchi, T.; Bernevig, B. A.; Yazdani, A. Cascade of Electronic Transitions in Magic-Angle Twisted Bilayer Graphene. *Nature* **2020**, *582* (7811), 198–202. https://doi.org/10.1038/s41586-020-2339-0.

(17) Stepanov, P.; Das, I.; Lu, X.; Fahimniya, A.; Watanabe, K.; Taniguchi, T.; Koppens, F. H. L.; Lischner, J.; Levitov, L.; Efetov, D. K. Untying the Insulating and Superconducting Orders in Magic-Angle Graphene. *Nature* **2020**, *583* (7816), 375–378. https://doi.org/10.1038/s41586-020-2459-6.

(18) Stepanov, P.; Xie, M.; Taniguchi, T.; Watanabe, K.; Lu, X.; MacDonald, A. H.; Bernevig, B. A.; Efetov, D. K. Competing Zero-Field Chern Insulators in Superconducting Twisted Bilayer Graphene. *Phys. Rev. Lett.* **2021**, *127* (19), 197701. https://doi.org/10.1103/PhysRevLett.127.197701.

(19) Guerci, D. Higher-Order Van Hove Singularity in Magic-Angle Twisted Trilayer Graphene. *Phys. Rev. Res.* **2022**, *4* (1). https://doi.org/10.1103/PhysRevResearch.4.L012013.

(20) Chen, G.; Jiang, L.; Wu, S.; Lyu, B.; Li, H.; Chittari, B. L.; Watanabe, K.; Taniguchi, T.; Shi, Z.; Jung, J.; Zhang, Y.; Wang, F. Evidence of a Gate-Tunable Mott Insulator in a Trilayer Graphene Moiré Superlattice. *Nat. Phys.* **2019**, *15* (3), 237–241. https://doi.org/10.1038/s41567-018-0387-2.

(21) Zhu, Z. Twisted Trilayer Graphene: A Precisely Tunable Platform for Correlated Electrons. *Phys. Rev. Lett.* **2020**, *125* (11). https://doi.org/10.1103/PhysRevLett.125.116404.

(22) Uri, A.; de la Barrera, S. C.; Randeria, M. T.; Rodan-Legrain, D.; Devakul, T.; Crowley, P. J. D.; Paul, N.; Watanabe, K.; Taniguchi, T.; Lifshitz, R.; Fu, L.; Ashoori, R. C.; Jarillo-Herrero, P. Superconductivity and Strong Interactions in a Tunable Moiré Quasicrystal. *Nature* **2023**, *620* (7975), 762–767. https://doi.org/10.1038/s41586-023-06294-z.

(23) Liang, M.; Xiao, M.-M.; Ma, Z.; Gao, J.-H. Moiré Band Structures of the Double Twisted Few-Layer Graphene. *Phys. Rev. B* **2022**, *105* (19), 195422. https://doi.org/10.1103/PhysRevB.105.195422.

(24) Gao, Q.; Khalaf, E. Symmetry Origin of Lattice Vibration Modes in Twisted Multilayer Graphene: Phasons versus Moiré Phonons. *Phys. Rev. B* **2022**, *106* (7), 075420. https://doi.org/10.1103/PhysRevB.106.075420.

(25) Chumak, A. V.; Vasyuchka, V. I.; Serga, A. A.; Hillebrands, B. Magnon Spintronics. *Nature Phys* **2015**, *11* (6), 453–461. https://doi.org/10.1038/nphys3347.

(26) Kruglyak, V. V.; Demokritov, S. O.; Grundler, D. Magnonics. *J. Phys. D: Appl. Phys.* **2010**, *43* (26), 264001. https://doi.org/10.1088/0022-3727/43/26/264001.
(27) Haldar, A.; Kumar, D.; Adeyeye, A. O. A Reconfigurable Waveguide for Energy-Efficient Transmission and Local Manipulation of Information in a Nanomagnetic Device. *Nature Nanotech* **2016**, *11* (5), 437–443. https://doi.org/10.1038/nnano.2015.332.
(28) Pirro, P.; Vasyuchka, V. I.; Serga, A. A.; Hillebrands, B. Advances in Coherent Magnonics. *Nat Rev Mater* **2021**, *6* (12), 1114–1135. https://doi.org/10.1038/s41578-021-00332-w.
(29) Khitun, A.; Bao, M.; Wang, K. L. Magnonic Logic Circuits. *J. Phys. D: Appl. Phys.* **2010**, *43* (26), 264005. https://doi.org/10.1088/0022-3727/43/26/264005.
(30) Csaba, G.; Papp, Á.; Porod, W. Perspectives of Using Spin Waves for Computing and Signal Processing. *Physics Letters A* **2017**, *381* (17), 1471–1476. https://doi.org/10.1016/j.physleta.2017.02.042.
(31) Talmelli, G.; Devolder, T.; Träger, N.; Förster, J.; Wintz, S.; Weigand, M.; Stoll, H.; Heyns, M.; Schütz, G.; Radu, I. P.; Gräfe, J.; Ciubotaru, F.; Adelmann, C. Reconfigurable Submicrometer Spin-Wave Majority Gate with Electrical Transducers. *Science Advances* **2020**, *6* (51), eabb4042. https://doi.org/10.1126/sciadv.abb4042.
(32) AENEAS; ARTEMIS-IA; EPoSS. *Electronic Components and Systems Strategic Research Agenda (ECS-SRA) 2020*; 2020. https://aeneas-office.org/2020/01/14/ecs-sra-2020-now-available-2/ (accessed 2025-11-26).
(33) Li, Y.-H.; Cheng, R. Moir\'e Magnons in Twisted Bilayer Magnets with Collinear Order. *Phys. Rev. B* **2020**, *102* (9), 094404. https://doi.org/10.1103/PhysRevB.102.094404.
(34) Ganguli, S. C.; Aapro, M.; Kezilebieke, S.; Amini, M.; Lado, J. L.; Liljeroth, P. Visualization of Moiré Magnons in Monolayer Ferromagnet. *Nano Lett.* **2023**, *23* (8), 3412–3417. https://doi.org/10.1021/acs.nanolett.3c00417.
(35) Li, M.; Jin, Z.; Zeng, Z.; Yan, P. Frequency Comb in Twisted Magnonic Crystals. arXiv July 15, 2025. https://doi.org/10.48550/arXiv.2507.10922.
(36) Ghader, D. Magnon Magic Angles and Tunable Hall Conductivity in 2D Twisted Ferromagnetic Bilayers. *Sci Rep* **2020**, *10* (1), 15069. https://doi.org/10.1038/s41598-020-72000-y.
(37) Flebus, B.; Grundler, D.; Rana, B.; Otani, Y.; Barsukov, I.; Barman, A.; Gubbiotti, G.; Landeros, P.; Akerman, J.; Ebels, U.; Pirro, P.; Demidov, V. E.; Schultheiss, K.; Csaba, G.; Wang, Q.; Ciubotaru, F.; Nikonov, D. E.; Che, P.; Hertel, R.; Ono, T.; Afanasiev, D.; Mentink, J.; Rasing, T.; Hillebrands, B.; Kusminskiy, S. V.; Zhang, W.; Du, C. R.; Finco, A.; van der Sar, T.; Luo, Y. K.; Shiota, Y.; Sklenar, J.; Yu, T.; Rao, J. The 2024 Magnonics Roadmap. *J. Phys.: Condens. Matter* **2024**, *36* (36), 363501. https://doi.org/10.1088/1361-648X/ad399c.
(38) Chen, J.; Zeng, L.; Wang, H.; Madami, M.; Gubbiotti, G.; Liu, S.; Zhang, J.; Wang, Z.; Jiang, W.; Zhang, Y.; Yu, D.; Ansermet, J.-P.; Yu, H. Magic-Angle Magnonic Nanocavity in a Magnetic Moir\'e Superlattice. *Phys. Rev. B* **2022**, *105* (9), 094445. https://doi.org/10.1103/PhysRevB.105.094445.
(39) Chen, J.; Madami, M.; Gubbiotti, G.; Yu, H. Magnon Confinement in a Nanomagnonic Waveguide by a Magnetic Moiré Superlattice. *Appl. Phys. Lett.* **2024**, *125* (16). https://doi.org/10.1063/5.0230523.
(40) Wang, H.; Madami, M.; Chen, J.; Jia, H.; Zhang, Y.; Yuan, R.; Wang, Y.; He, W.; Sheng, L.; Zhang, Y.; Wang, J.; Liu, S.; Shen, K.; Yu, G.; Han, X.; Yu, D.; Ansermet, J.-P.; Gubbiotti, G.; Yu, H. Observation of Spin-Wave Moir\'e Edge and Cavity Modes in Twisted Magnetic Lattices. *Phys. Rev. X* **2023**, *13* (2), 021016. https://doi.org/10.1103/PhysRevX.13.021016.
(41) Xi, Y.; Shi, Z.; Zhao, M.; Cheng, N.; Du, K.; Li, K.; Xu, H.; Xu, S.; Liu, J.; Feng, H.; Shi, Y.; Xu, X.; Hao, W.; Dou, S.; Du, Y. Modulation of Kondo Behavior in a Two-Dimensional Epitaxial Bilayer Bi(111)/Fe3GeTe2 Moiré Heterostructure. *ACS Nano* **2024**, *18* (34), 22958–22964. https://doi.org/10.1021/acsnano.4c04271.
(42) Xu, Y.; Ray, A.; Shao, Y.-T.; Jiang, S.; Lee, K.; Weber, D.; Goldberger, J. E.; Watanabe, K.; Taniguchi, T.; Muller, D. A.; Mak, K. F.; Shan, J. Coexisting Ferromagnetic–Antiferromagnetic State in Twisted Bilayer CrI3. *Nat. Nanotechnol.* **2022**, *17* (2), 143–147. https://doi.org/10.1038/s41565-021-01014-y.
(43) Wang, Z.; Rhodes, D. A.; Watanabe, K.; Taniguchi, T.; Hone, J. C.; Shan, J.; Mak, K. F. Evidence of High-Temperature Exciton Condensation in Two-Dimensional Atomic Double Layers. *Nature* **2019**, *574* (7776), 76–80. https://doi.org/10.1038/s41586-019-1591-7.

(44) Koperski, M.; Nogajewski, K.; Arora, A.; Cherkez, V.; Mallet, P.; Veuillen, J.-Y.; Marcus, J.; Kossacki, P.; Potemski, M. Single Photon Emitters in Exfoliated WSe2 Structures. *Nature Nanotech* **2015**, *10* (6), 503–506. https://doi.org/10.1038/nnano.2015.67.

(45) Wang, C.; Gao, Y.; Lv, H.; Xu, X.; Xiao, D. Stacking Domain Wall Magnons in Twisted van Der Waals Magnets. *Phys. Rev. Lett.* **2020**, *125* (24), 247201. https://doi.org/10.1103/PhysRevLett.125.247201.

(46) Gubbiotti, G.; Sadovnikov, A.; Beginin, E.; Nikitov, S.; Wan, D.; Gupta, A.; Kundu, S.; Talmelli, G.; Carpenter, R.; Asselberghs, I.; Radu, I. P.; Adelmann, C.; Ciubotaru, F. Magnonic Band Structure in Vertical Meander-Shaped ${\mathrm{Co}}_{40}${\mathrm{Fe}}_{40}${\mathrm{B}}_{20}$ Thin Films. *Phys. Rev. Appl.* **2021**, *15* (1), 014061. https://doi.org/10.1103/PhysRevApplied.15.014061.

(47) Martyshkin, A. A.; Beginin, E. N.; Stognij, A. I.; Nikitov, S. A.; Sadovnikov, A. V. Vertical Spin-Wave Transport in Magnonic Waveguides With Broken Translation Symmetry. *IEEE Magnetics Letters* **2019**, *10*, 1–5. https://doi.org/10.1109/LMAG.2019.2957264.

(48) Lin, F.; Qiao, J.; Huang, J.; Liu, J.; Fu, D.; Mayorov, A. S.; Chen, H.; Mukherjee, P.; Qu, T.; Sow, C.-H.; Watanabe, K.; Taniguchi, T.; Özyilmaz, B. Heteromoiré Engineering on Magnetic Bloch Transport in Twisted Graphene Superlattices. *Nano Lett.* **2020**, *20* (10), 7572–7579. https://doi.org/10.1021/acs.nanolett.0c03062.

(49) Zhang, X.; Tsai, K.-T.; Zhu, Z.; Ren, W.; Luo, Y.; Carr, S.; Luskin, M.; Kaxiras, E.; Wang, K. Correlated Insulating States and Transport Signature of Superconductivity in Twisted Trilayer Graphene Superlattices. *Phys. Rev. Lett.* **2021**, *127* (16), 166802. https://doi.org/10.1103/PhysRevLett.127.166802.

(50) Yu, H.; d'Allivy Kelly, O.; Cros, V.; Bernard, R.; Bortolotti, P.; Anane, A.; Brandl, F.; Huber, R.; Stasinopoulos, I.; Grundler, D. Magnetic Thin-Film Insulator with Ultra-Low Spin Wave Damping for Coherent Nanomagnonics. *Sci Rep* **2014**, *4* (1), 6848. https://doi.org/10.1038/srep06848.

(51) Chang, H.; Li, P.; Zhang, W.; Liu, T.; Hoffmann, A.; Deng, L.; Wu, M. Nanometer-Thick Yttrium Iron Garnet Films With Extremely Low Damping. *IEEE Magnetics Letters* **2014**, *5*, 1–4. https://doi.org/10.1109/LMAG.2014.2350958.

(52) Kajiwara, Y.; Harii, K.; Takahashi, S.; Ohe, J.; Uchida, K.; Mizuguchi, M.; Umezawa, H.; Kawai, H.; Ando, K.; Takanashi, K.; Maekawa, S.; Saitoh, E. Transmission of Electrical Signals by Spin-Wave Interconversion in a Magnetic Insulator. *Nature* **2010**, *464* (7286), 262–266. https://doi.org/10.1038/nature08876.

(53) Neusser, S.; Duerr, G.; Bauer, H. G.; Tacchi, S.; Madami, M.; Woltersdorf, G.; Gubbiotti, G.; Back, C. H.; Grundler, D. Anisotropic Propagation and Damping of Spin Waves in a Nanopatterned Antidot Lattice. *Phys. Rev. Lett.* **2010**, *105* (6), 067208. https://doi.org/10.1103/PhysRevLett.105.067208.

(54) Groß, F.; Zelent, M.; Gangwar, A.; Mamica, S.; Gruszecki, P.; Werner, M.; Schütz, G.; Weigand, M.; Goering, E. J.; Back, C. H.; Krawczyk, M.; Gräfe, J. Phase Resolved Observation of Spin Wave Modes in Antidot Lattices. *Appl. Phys. Lett.* **2021**, *118* (23). https://doi.org/10.1063/5.0045142.

(55) Lou, B.; Zhao, N.; Minkov, M.; Guo, C.; Orenstein, M.; Fan, S. Theory for Twisted Bilayer Photonic Crystal Slabs. *Phys. Rev. Lett.* **2021**, *126* (13), 136101. https://doi.org/10.1103/PhysRevLett.126.136101.

(56) Eshbach, J. R.; Damon, R. W. Surface Magnetostatic Modes and Surface Spin Waves. *Phys. Rev.* **1960**, *118* (5), 1208–1210. https://doi.org/10.1103/PhysRev.118.1208.

(57) Yamamoto, K.; Thiang, G. C.; Pirro, P.; Kim, K.-W.; Everschor-Sitte, K.; Saitoh, E. Topological Characterization of Classical Waves: The Topological Origin of Magnetostatic Surface Spin Waves. *Phys. Rev. Lett.* **2019**, *122* (21), 217201. https://doi.org/10.1103/PhysRevLett.122.217201.

(58) Dong, K.; Zhang, T.; Li, J.; Wang, Q.; Yang, F.; Rho, Y.; Wang, D.; Grigoropoulos, C. P.; Wu, J.; Yao, J. Flat Bands in Magic-Angle Bilayer Photonic Crystals at Small Twists. *Phys. Rev. Lett.* **2021**, *126* (22), 223601. https://doi.org/10.1103/PhysRevLett.126.223601.

(59) Lopes dos Santos, J. M. B.; Peres, N. M. R.; Castro Neto, A. H. Graphene Bilayer with a Twist: Electronic Structure. *Phys. Rev. Lett.* **2007**, *99* (25), 256802. https://doi.org/10.1103/PhysRevLett.99.256802.

(60) Ramires, A.; Lado, J. L. Emulating Heavy Fermions in Twisted Trilayer Graphene. *Phys. Rev. Lett.* **2021**, *127* (2), 026401. https://doi.org/10.1103/PhysRevLett.127.026401.

(61) Christos, M.; Sachdev, S.; Scheurer, M. S. Correlated Insulators, Semimetals, and Superconductivity in Twisted Trilayer Graphene. *Phys. Rev. X* **2022**, *12* (2), 021018. https://doi.org/10.1103/PhysRevX.12.021018.

(62) Zhang, Y.; Polski, R.; Lewandowski, C.; Thomson, A.; Peng, Y.; Choi, Y.; Kim, H.; Watanabe, K.; Taniguchi, T.; Alicea, J.; Von Oppen, F.; Refael, G.; Nadj-Perge, S. Promotion of Superconductivity in Magic-Angle Graphene Multilayers. *Science* **2022**, *377* (6614), 1538–1543. https://doi.org/10.1126/science.abn8585.

(63) Hao, Z.; Zimmerman, A. M.; Ledwith, P.; Khalaf, E.; Najafabadi, D. H.; Watanabe, K.; Taniguchi, T.; Vishwanath, A.; Kim, P. Electric Field–Tunable Superconductivity in Alternating-Twist Magic-Angle Trilayer Graphene. *Science* **2021**, *371* (6534), 1133–1138. https://doi.org/10.1126/science.abg0399.

(64) Park, J. M.; Cao, Y.; Watanabe, K.; Taniguchi, T.; Jarillo-Herrero, P. Tunable Strongly Coupled Superconductivity in Magic-Angle Twisted Trilayer Graphene. *Nature* **2021**, *590* (7845), 249–255. https://doi.org/10.1038/s41586-021-03192-0.

(65) Kim, H.; Choi, Y.; Lewandowski, C.; Thomson, A.; Zhang, Y.; Polski, R.; Watanabe, K.; Taniguchi, T.; Alicea, J.; Nadj-Perge, S. Evidence for Unconventional Superconductivity in Twisted Trilayer Graphene. *Nature* **2022**, *606* (7914), 494–500. https://doi.org/10.1038/s41586-022-04715-z.

(66) Xu, T.; Yang, S.; Nair, S. V.; Ruda, H. E. Confined Modes in Finite-Size Photonic Crystals. *Phys. Rev. B* **2005**, *72* (4), 045126. https://doi.org/10.1103/PhysRevB.72.045126.

(67) Zhu, Z.; Cazeaux, P.; Luskin, M.; Kaxiras, E. Modeling Mechanical Relaxation in Incommensurate Trilayer van Der Waals Heterostructures. *Phys. Rev. B* **2020**, *101* (22), 224107. https://doi.org/10.1103/PhysRevB.101.224107.

(68) Vansteenkiste, A.; Leliaert, J.; Dvornik, M.; Helsen, M.; Garcia-Sanchez, F.; Van Waeyenberge, B. The Design and Verification of MuMax3. *AIP Advances* **2014**, *4* (10), 107133. https://doi.org/10.1063/1.4899186.